\documentclass[pra,aps,twocolumn,superscriptaddress,showpacs,floatfix]{revtex4}
\usepackage{amsmath}

\usepackage{graphicx}

\def\prn#1{{\left(#1\right)}}

\def\threej(#1,#2)(#3,#4)(#5,#6){\begin{pmatrix}#1&#3&#5\\#2&#4&#6\end{pmatrix}}
\def\sixj(#1,#2,#3)(#4,#5,#6){\begin{Bmatrix}#1&#2&#3\\#4&#5&#6\end{Bmatrix}}
\def\ninej(#1,#2,#3)(#4,#5,#6)(#7,#8,#9){\begin{Bmatrix}#1&#2&#3\\#4&#5&#6\\#7&#8&#9\end{Bmatrix}}

\begin{document}

\setkeys{Gin}{width=3.25 in}

\title{Investigation of microwave transitions and nonlinear magneto-optical rotation in anti-relaxation-coated cells\footnote{This
work is a partial contribution of NIST, an agency of the US
Government, and is not subject to copyright.}}
\author{D. Budker}\email{budker@socrates.berkeley.edu}
\affiliation{Department of Physics, University of California at
Berkeley, Berkeley, California 94720-7300} \affiliation{Nuclear
Science Division, Lawrence Berkeley National Laboratory, Berkeley,
California 94720}
\author{L. Hollberg}
\affiliation{National Institute
of Standards and Technology, 325 S. Broadway, Boulder, CO
80305-3322}
\author{D. F. Kimball}
\affiliation{Department of Physics, University of California at
Berkeley, Berkeley, California 94720-7300}
\author{J. Kitching}\email{kitching@boulder.nist.gov}
\affiliation{National Institute of Standards and Technology, 325
S. Broadway, Boulder, CO 80305-3322}
\author{S. Pustelny}\affiliation{Instytut Fizyki im. M.
Smoluchowskiego, Uniwersytet Jagiello\'{n}ski, Reymonta 4, 30-059
Krakow, Poland}
\author{V. V. Yashchuk}
\affiliation{Advanced Light Source Division, Lawrence Berkeley
National Laboratory, Berkeley CA 94720}

\date{\today}

\begin{abstract}
Using laser optical pumping, widths and frequency shifts are
determined for microwave transitions between ground-state
hyperfine components of $^{85}$Rb and $^{87}$Rb atoms contained in
vapor cells with alkane anti-relaxation coatings. The results are
compared with data on Zeeman relaxation obtained in nonlinear
magneto-optical rotation (NMOR) experiments, a comparison
important for quantitative understanding of spin-relaxation
mechanisms in coated cells. By comparing cells manufactured over a
forty-year period we demonstrate the long-term stability of coated
cells, an important property for atomic clocks and magnetometers.
\end{abstract}



\maketitle

\section{Introduction}

Alkali-metal vapor cells with paraffin anti-relaxation coatings,
first introduced by H. G. Robinson et al. in the 1950's
\cite{Rob58} and subsequently studied in great detail by M.-A.
Bouchiat et. al. \cite{Bou64,BouBro66} and other authors (see, for
example, Ref. \cite{Lib86} and references therein), are presently
at the heart of some of the most sensitive optical-pumping
magnetometers \cite{Ale92}. Recently, paraffin-coated cells have
also been used in investigations of nonlinear-optics such as
nonlinear magneto- and electro-optical effects (see Refs.
\cite{Bud2002RMP,Ale2005} for detailed reviews). Using coated
cells, Zeeman and nonlinear-optical resonance widths of less than
1 Hz and microwave resonance widths of several hertz have been
achieved. Although the use of paraffin-coated cells in atomic
clocks has also been extensively investigated (see, for example,
Refs. \cite{Ris80,Rob82,Fru83}), up until now such cells have not
been commonly used in commercial clocks despite of their potential
advantages, including excellent short-term stability and
the sensitivity of the resonance frequency to temperature
variations comparable to that of buffer-gas cells (when the gas
composition in the case of buffer-gas cells is chosen to minimize
the temperature dependence).

The ongoing work aimed at developing highly miniaturized atomic
frequency references \cite{Kit2002,Jau2004} and magnetometers has
created renewed interest in wall coatings. These devices will
likely take advantage of miniature atomic vapor cells with
physical dimensions on the order of $1\ $mm or smaller
\cite{Lie2004}. Because of the larger surface-to-volume ratio,
atoms confined in such a small cell spend a larger fraction of
their time interacting with the cell wall than they would in a
larger cell. Therefore, linewidth broadening and frequency shifts
associated with the cell walls can significantly affect the
performance of the frequency reference even when a buffer gas of
moderate pressure is used. Thus the quality and efficiency of
coatings applied to the walls of such small cells is an important
factor in determining the feasibility of highly miniaturized
frequency references, particularly those with cell volumes
significantly below $1\ $mm$^3$.

Collisions of alkali atoms with cell walls can affect the atomic
state in a number of ways. In addition to optical resonance
broadening, the atomic ground states can undergo Zeeman and
hyperfine decoherence and population transfer. Typically, the
hyperfine decoherence rate is found to be about one order of
magnitude larger than the Zeeman decoherence rate
\cite{Rob82,Rob83}. For atomic frequency references, the
properties of the atomic transition between hyperfine levels
determine the short-term stability, while for most magnetometers,
transitions between Zeeman levels within a single hyperfine level
are important. Thus, investigation of a comparison between
different wall-relaxation effects, particularly a comparison of
Zeeman and hyperfine decoherence, is of interest both from the
viewpoint of understanding the basic physics and with regard to
practical applications.

In this paper, we present measurements of intrinsic linewidths and
shifts performed with several alkane-coated alkali-vapor cells
manufactured in different laboratories over a period of about 40
years. We compare the results to those of the nonlinear
magneto-optical rotation (NMOR \cite{Bud2002RMP}) experiments, and
draw conclusions pertaining to the applications of the coated
cells to atomic clocks. Specifically, we find that wall coatings
can remain effective for at least several decades after the cell
is fabricated, and that for high-quality alkane coatings, neither
the molecular weight distribution of the alkane chains nor the
method by which the wall coating is deposited has much bearing on
the effectiveness of the coating. In addition, the NMOR linewidth
is found to be up to ten times smaller than the hyperfine
coherence linewidth when the two are measured in the same cell at
similar temperatures. This can be explained by the way in which
electron-spin randomization contributes to the decoherence in each
of these cases.

\section{The cells}
The anti-relaxation-coated buffer-gas-free cells used in the
present work are listed in Table \ref{Table1}. The inner-wall
coating of each of these spherical cells was an alkane film
(general chemical formula $\rm{C_{n}H_{2n+2}}$). None of these
cells were ``re-cured" or heat treated before testing (a procedure
that may be used, if necessary, to reduce the relaxation due to
alkali metal that has collected on the coated walls).

The cell identified in Table \ref{Table1} as Gib, containing a
natural mixture of Rb isotopes, was manufactured at Berkeley
around 1964 for the Ph.D. thesis work of H. M. Gibbs. (The Thesis
\cite{Gib65} gives a detailed description of the cell
manufacturing process.) The coating material is Paraflint, which
consists of alkane chains with a wide range of molecular weights.
The coating was applied by melting Paraflint wax and running it
over the cell surface. This cell has obvious Rb crystalline
deposits on the wall and there are several glass tubulations on
the spherical surface (i.e., imperfections that originate from
glass tubes that were attached to the cell during the
manufacturing process). Regions of the cell have heavy pooling of
wax and the coating thickness is very non-uniform.

The $^{85}$Rb cell referred to as Ale-10 was made by E. B.
Alexandrov and M. V. Balabas according to the procedure described,
for example, in Ref. \cite{AleLIAD}. The material of the coating
is a mixture of alkane chains of different length (n$\sim$50)
fractionated from polyethylene at 220$^{\circ}$C. The coating was
applied by vapor deposition, where a piece of the paraffin was
placed in a side-arm and the whole cell was heated to fill it with
the paraffin vapor, and then cooled. The resulting coating has
estimated thickness of $\sim 10\ \mu$m, and can be barely seen by
eye.

The $^{87}$Rb cells referred to as TT11 and H2 were made by H. G.
Robinson \cite{Rob85unp}. The material of the coating is
tetracontane (n=40) distilled at about 200$^{\circ}$C. The
purified tetracontane wax was evaporatively coated onto the cell
surface from a hot needle. The resulting coating is thin (and so
not visible by eye) and has a melting temperature of $\sim
80^{\circ}$C.

Each cell with the exception of Gib was made with an uncoated stem
containing rubidium metal (see also Section
\ref{Sect:Interpretation}).

\section{Microwave-transition measurements}
\begin{figure}
\includegraphics{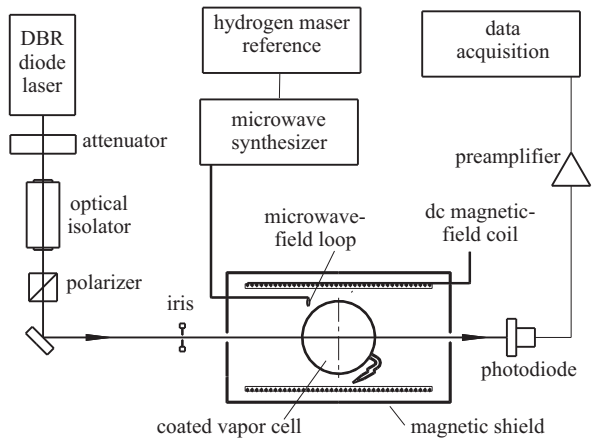}
\caption{Experimental setup for measuring microwave transitions.}
\label{FigApparatus}
\end{figure}

A schematic of the apparatus is shown in Fig.\ref{FigApparatus}.
We used a
distributed-Bragg-reflector (DBR) diode laser \cite{Hir91} with a
measured linewidth of $\approx 2\ $MHz, whose frequency was tuned
near the Rb D1 resonance ($\lambda=795\ $nm). The light beam
passes through a variable attenuator, an optical isolator used to
avoid optical feedback into the laser, and then a linear
polarizer. It is then directed through a vapor cell (at room
temperature) enclosed in a single-layer cylindrical magnetic
shield. The laser beam diameter at the location of the cell is
$\sim 7\ $mm. The intensity of the transmitted light is detected
with a photodiode. A dc magnetic field is applied to the cell
parallel to the direction of light propagation by passing current
through a coil wound on the surface of a cylindrical acrylic
insert (not shown) that fits into the magnetic shield. A microwave
field is applied with a single-wire loop terminating a coaxial
cable. The microwave synthesizer is referenced to a hydrogen maser
($\delta \nu/\nu\approx 2\times10^{-13}\tau^{-1/2}$, where $\tau$
is measured in seconds).

For one cell (TT11 \footnote{This cell was kindly loaned to us by
Symmetricom, TRC. This does not imply an endorsement by NIST;
cells from other companies may work equally well.}, see Table
\ref{Table1}), a slightly different apparatus was used. The
magnetic shield consisted of two layers, and a low-Q (of several
thousand) cylindrical microwave cavity operating in the TE$_{011}$
mode was employed instead of the wire loop. In this apparatus, it
was possible to heat the cell while keeping the stem at a lower
temperature. This was useful for evaluating the effect of the
temperature on the widths and shifts of the microwave transition.

\begin{figure}[!h]
\includegraphics{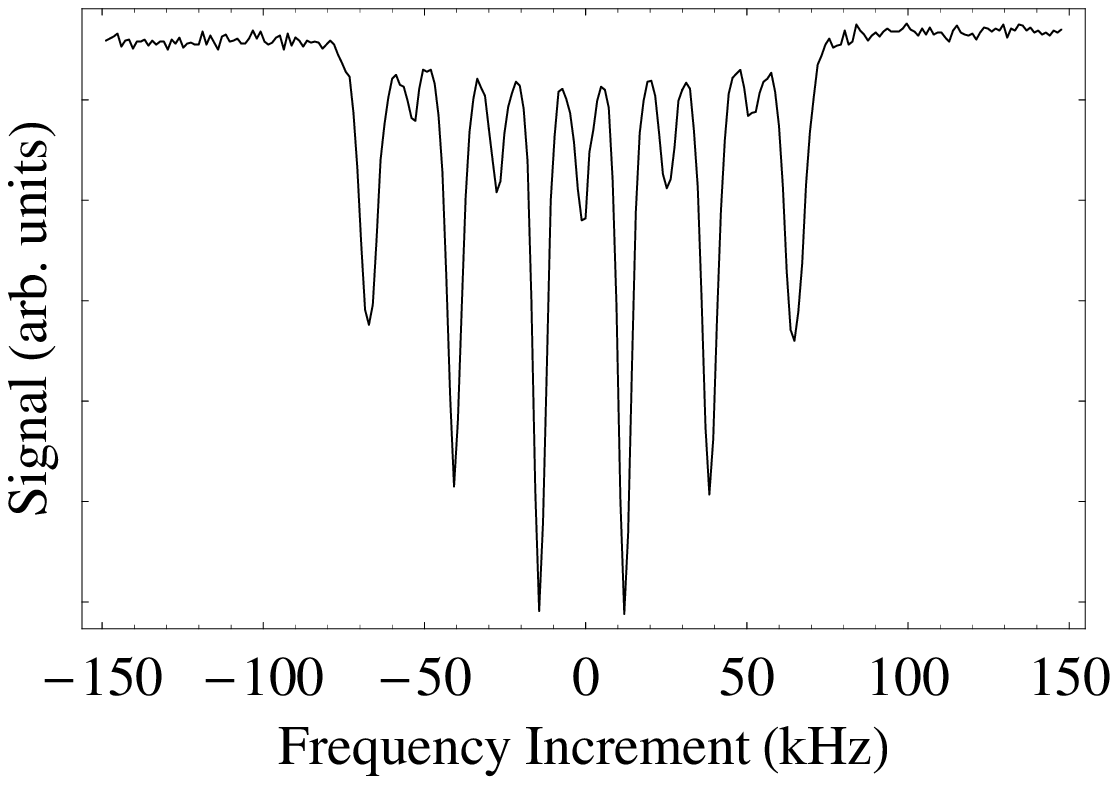}
\includegraphics{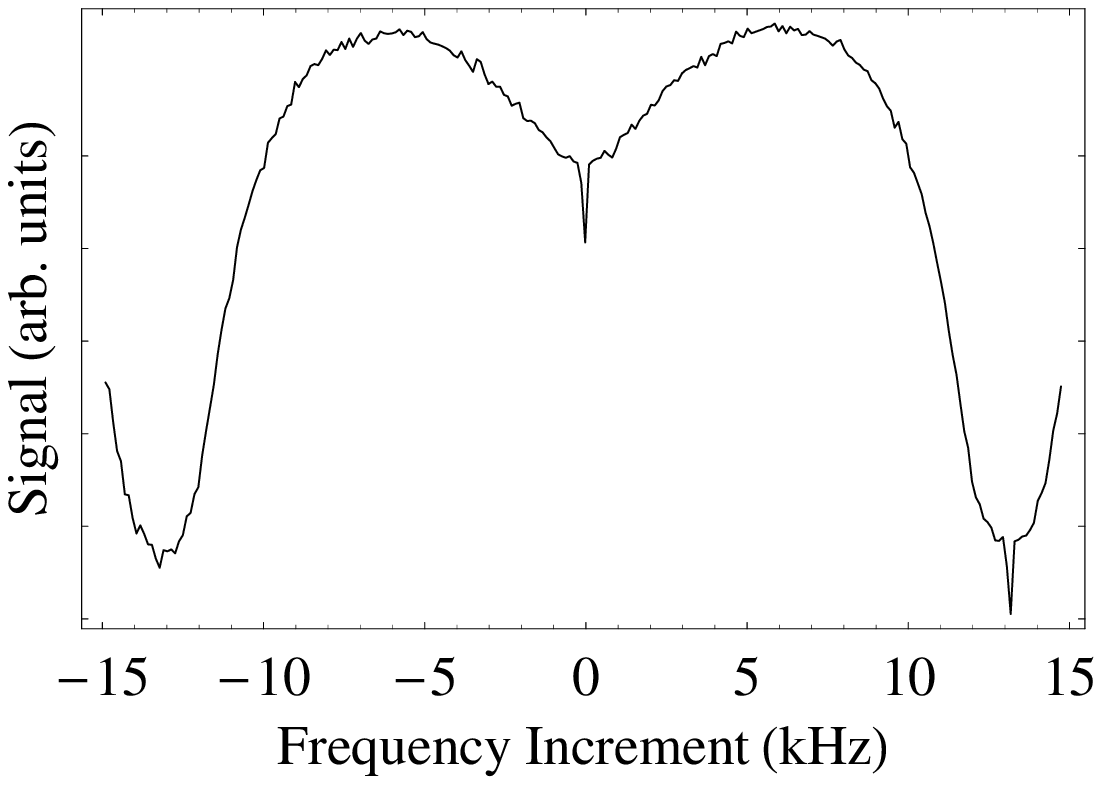}
\includegraphics{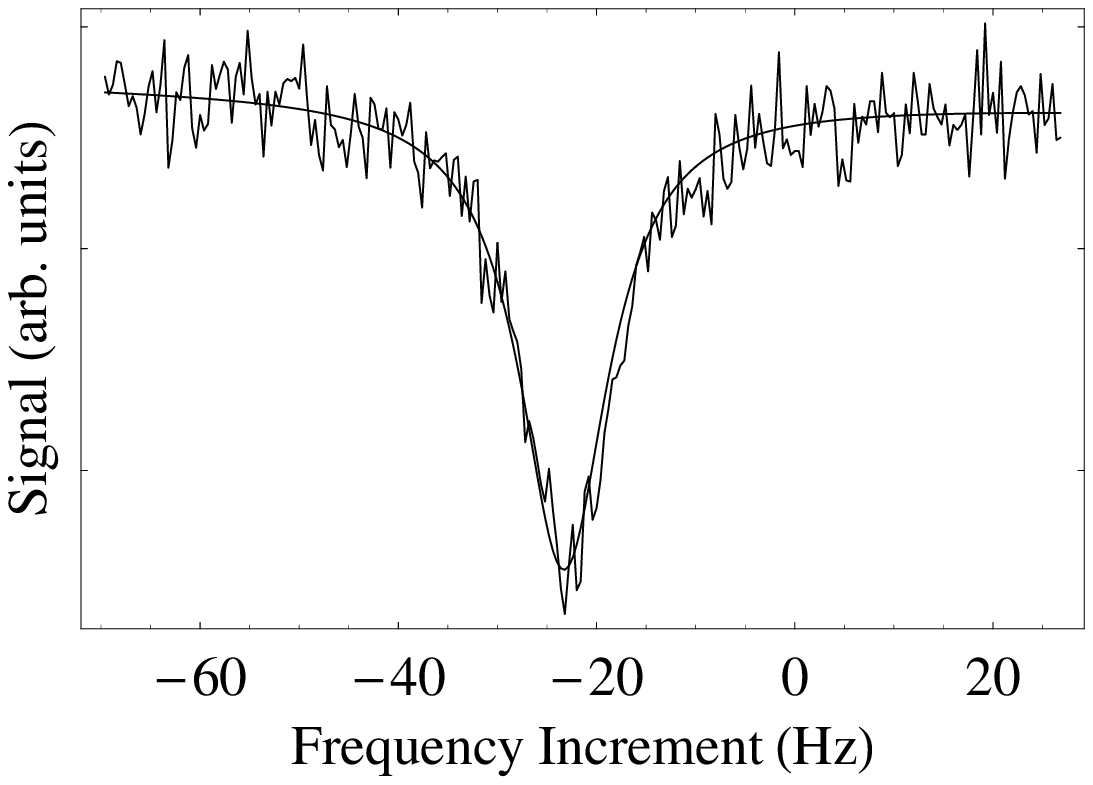}
\caption{Examples of the microwave spectra recorded with the
Ale-10 $^{85}$Rb cell (see Table \ref{Table1}). The signal
corresponds to the intensity of the laser light transmitted
through the cell. The common parameters for the three scans are:
dc magnetic field $=29\ $mG, linear light polarization, laser
tuned to the center of the $F=3\rightarrow F'$ transition, total
scan rate (saw-tooth sweep) $=0.02\ $Hz. Each plot represents an
average of approximately five scans. Upper and middle plots: input
light power $0.25\ \mu$W; lower plot: input light power $13\
\mu$W, microwave power reduced by a factor $\approx 63$. The lower
plot also shows a fit by a Lorentzian superimposed on a linear
background. The fit Lorentzian linewidth (FWHM) is $10.9(3)\ $Hz,
slightly larger than the ``intrinsic" width of about $8.7\ $Hz
(Table \ref{Table1}) due to residual light broadening (see text).}
\label{NISTuW2}
\end{figure}

\begin{figure}[!h]
\includegraphics{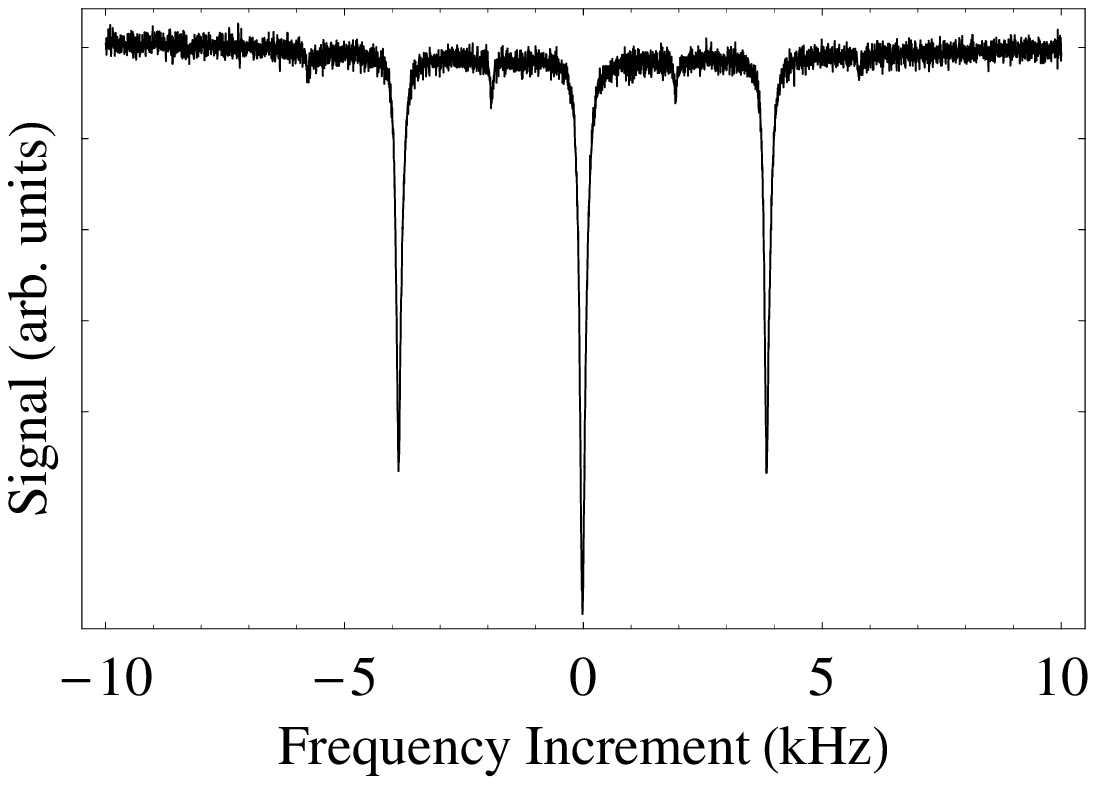}
\caption{An example of a microwave spectrum recorded with the TT11
$^{87}$Rb cell (see Table \ref{Table1}). The parameters for the
scan are: dc magnetic field $=2.7\ $mG, linear light polarization,
laser tuned to the center of the $F=2\rightarrow F'$ transition,
total scan rate (saw-tooth sweep) $=0.2\ $Hz. The plot represents
a single scan. Input light power $=6\ \mu$W. The comparison of
this plot with the middle plot in Fig. \ref{NISTuW2} illustrates
the advantages of using a TE$_{011}$ microwave cavity as opposed
to a current loop: almost complete suppression of the broad
pedestal and the $\Delta M_F \ne 0$ microwave transitions.}
\label{NISTuW5}
\end{figure}

In the experiment, the laser frequency was tuned to a particular
location on the optical absorption profile, and the microwave
frequency was swept around the nominal frequency of the field-free
separation of the ground-state hyperfine components of $^{85}$Rb:
$3,035,732,440\ $Hz \cite{Tet76} or $^{87}$Rb: $6,834,682,610.9\
$Hz \cite{Biz99}, while the transmitted light intensity was
recorded. An applied dc magnetic field gave rise to the splittings
as shown in the microwave spectra of Figs. \ref{NISTuW2} and
\ref{NISTuW5}.
\begin{table*}
 \caption{Vapor cells used for the measurements and
the experimentally determined values of microwave linewidths
($\gamma_{exp}^{\mu}$) and shifts. $T$ - the temperature of the
cell (which was higher than that for the stem for some of the
measurements). n - the total number density of Rb vapor. The
quantity $\phi$ is the estimated average phase shift per wall
collision. The uncertainty in $\phi$ includes an estimate of the
effect of non-ideal cell shape. For $^{87}$Rb, the phase shift
$\phi'$ scaled to $^{85}$Rb (see text) is also given. The last two
columns list the deduced contributions to the microwave linewidths
from the spin-exchange collisions and the adiabatic collisions,
respectively (see text).}\label{Table1}
   \begin{tabular}{c|c|c|c|c|c|c|c|c|c|c|c|c}
\hline
        Cell & Year made & Ref. & Diam. (cm) & $T$ ($^{\circ}$C)         & n (cm$^{-3}$)       & Isotope  & $\frac{\gamma_{exp}^{\mu}}{2\pi}$ (Hz) & Shift (Hz) & $|\phi|$ (rad)  & $|\phi'|$ (rad)  & $\frac{\gamma^{\mu}_{se}}{2\pi}$ (Hz) & $\frac{\gamma_a^{\mu}}{2\pi}$ (Hz)\\
        \hline \hline
         Ale-10 & 1997 & \cite{AleLIAD} & 10 & $25$ & $8\cdot 10^9$ & $^{85}$Rb  & 8.7(5)  & -24(4) & 0.037(7) & & 1.3 & 2 \\
          \hline
         Gib & $1964$ & \cite{Gib65} & 10 & $25$ & $8\cdot 10^9$ & $^{85}$Rb  & 11(2)  & -14(4) & 0.022(6) & & 1.3 & 0.6 \\
           &   &   &   &    &  & $^{87}$Rb  & 16(4) & -42(2) & 0.065(6) & 0.029(3) & 1.2 & 5 \\
          \hline
         H2 & 1985 & \cite{Rob85unp} & 3.5 & 21 & $7 \cdot 10^9$ & $^{87}$Rb  & 22(3)  & -93(1) & 0.050(5) & 0.022(2) & 1.2 & 9 \\
          \hline
         TT11 & 1985 & \cite{Rob85unp} & 3.4 & 22 & $6.5 \cdot 10^9$ & $^{87}$Rb  & 23(2)  & -80(1) & 0.043(4) & 0.019(2) & 1.0 & 7\\
          & & &  & 42  & $1.4\cdot 10^{10}$ & & 17.5(10)  &   &  &  & 2.2 & \\
          & & &  & 43  & $3.4\cdot 10^{10}$ & & 21.5(10)  & -70.5(3)  & 0.036(4) & 0.016(2) & 5.2 & 5\\
          \hline
          \hline
    \end{tabular}

\end{table*}
\begin{table*}
 \caption{Experimentally determined values for FM NMOR linewidths. $T$ - the temperature of the cell (the stem and the cell
body are at the same temperature). n - the total number density of
Rb vapor. $\gamma^{NMOR}_{se}$ - calculated spin-exchange
contribution to the NMOR linewidth. The last column lists the
deduced contributions to the microwave linewidths from
electron-spin-randomization collisions (see text).} \label{Table2}
   \begin{tabular}{c|c|c|c|c|c|c}
\hline
        Cell & $T$ ($^{\circ}$C)         & n (cm$^{-3}$) & Isotope  & $\gamma_{exp}^{NMOR}/(2\pi)$ (Hz) & $\gamma_{se}^{NMOR}/(2\pi)$ (Hz) & $\gamma_{er}^{\mu}/(2\pi)$ (Hz)\\
        \hline \hline
         Ale-10 & 19 & $4 \cdot 10^9$ & $^{85}$Rb & 0.7(1) & 0.15 & 3 \\
                & 25 & $6 \cdot 10^9$ &           & 1.2(1) & 0.24 & 5 \\
          \hline
         Gib & 21 & $4 \cdot 10^9$ & $^{85}$Rb & 2.9(1) & 0.15 & 13 \\
          &  &  & $^{87}$Rb & 2.9(1) & 0.24 & 9 \\
          \hline
         H2 & 21 & $4 \cdot 10^9$ & $^{87}$Rb  & 3.5(1) & 0.24 & 11\\
          \hline
          \hline
    \end{tabular}

\end{table*}

\section{Nonlinear magneto-optical measurements of Zeeman relaxation}

\begin{figure}
\includegraphics{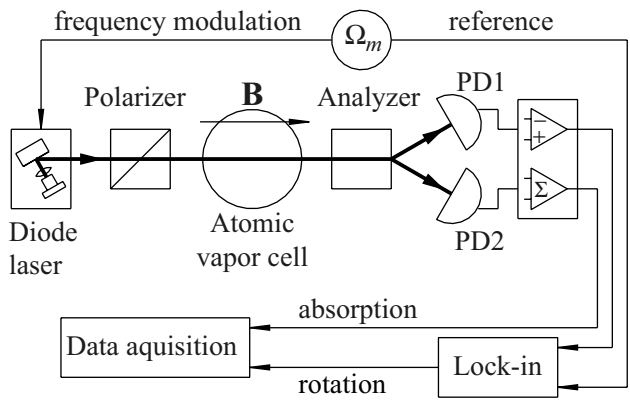}
\caption{Simplified schematic of the experimental setup for
measuring Zeeman relaxation rate with the FM NMOR technique
\cite{Bud2002FM,Yas2003Select,Mal2004}.}
\label{Fig_FMNMOR_Apparatus}
\end{figure}

For all cells except TT11, in addition to studying microwave
transitions, we also measured Zeeman relaxation rates using
nonlinear magneto-optical rotation. The general idea of the method
is the following (see the review \cite{Bud2002RMP} for a detailed
discussion). The interaction of the atoms with a near-resonant
laser field results in polarization of the atomic medium. The
induced polarization evolves in the presence of a magnetic field;
the resulting change of the medium's polarization is detected by
measuring optical rotation of the linearly polarized light (which,
in this case, plays the part of both the pump and the probe). In
this work, we used a version of the NMOR technique in which the
frequency of the laser is modulated and optical rotation varying
at the modulation frequency is detected (FM NMOR
\cite{Bud2002FM,Yas2003Select,Mal2004}; Fig.
\ref{Fig_FMNMOR_Apparatus}).

The vapor cell under study was placed inside a multi-layer
magnetic shielding system (not shown) equipped with coils for
compensating residual magnetic fields and gradients and for
applying well controlled fields to the cell. The time-dependent
optical rotation was detected with a balanced polarimeter after
the vapor cell. In this work, the laser was tuned to the Rb D1
line; the laser beam diameter was $\sim 2\ $mm, and the light
power was $\lesssim 15\ \mu$W. A static magnetic field was applied
along the light propagation direction. Narrow resonances in
modulation frequency appear in the synchronously detected
optical-rotation signal when the frequency is equal to twice the
Larmor frequency (Fig. \ref{Fig_FMNMOR_Data}). The factor of two
is related to the two-fold spatial symmetry of the induced atomic
alignment. The widths of the resonances are determined by the
ground-state Zeeman relaxation rate.

An advantage of the FM NMOR technique for measuring the Zeeman
relaxation rate is that when a bias field is applied (as in the
present case), the resonance curves are, to first order,
insensitive to small transverse magnetic fields; this eliminates a
possible source of systematic error.

\section{Procedure, results and discussion}

The total electronic angular momentum in the ground electronic
state of rubidium is $J=1/2$. For $^{85}$Rb, the nuclear spin is
$I=5/2$. In a low dc magnetic field where nonlinear Zeeman shifts
resulting from decoupling of hyperfine structure can be neglected,
one generally expects to see 11 distinct resonance hyperfine
transition frequencies between various linear-Zeeman-split
sublevels of the upper and lower ground-state hyperfine levels.
(For $^{87}$Rb, an atom with $I=3/2$, there are 7 transition
frequencies.) The experimental observation of this is shown in the
upper plot of Fig. \ref{NISTuW2}).  The relative intensities of
various components depend on the power, polarization, and tuning
of the pump light, the geometry and orientation of the microwave
loop, and the microwave power. Under typical conditions in this
experiment, the peaks of the resonances correspond to an increase
in absorption by several percent (the optical depth on resonance
of the room-temperature cells is of order unity).

The width of the peaks on the upper plot in Fig. \ref{NISTuW2} is
about $4\ $kHz (FWHM) and is dominated by the Doppler width of the
microwave transition. There is also a significant contribution due
to the phase variation of the microwave field over the cell volume
resulting from the use of a simple loop for generation of the
microwave field. A zoom with higher frequency resolution (the
middle plot in Fig. \ref{NISTuW2}) reveals additional sharp
features superimposed on top of the Doppler-broadened lines. These
are the Dicke-narrowed lines \cite{Dic53,Rob82} of primary
interest in the present work.

While all the sharp features are of comparable widths, in the
following we concentrate on the central resonance corresponding to
the ``clock transition" between the $M=0$ Zeeman components (the
0-0 transition), which is to first order insensitive to dc
magnetic fields and gradients.
\begin{figure}
\includegraphics{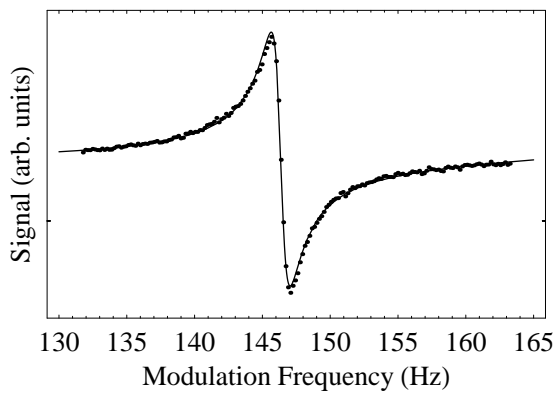}
\caption{An example of the FM NMOR data (the lock-in detector
output, see Fig. \ref{Fig_FMNMOR_Apparatus}) taken with the Ale-10
cell. The central frequency of the laser is tuned $\approx 400\
$MHz lower than the center of the $F=3\rightarrow F'$ absorption
peak of the D1 line (this point corresponds to a maximum of the FM
NMOR signal). The laser frequency is modulated with an amplitude
of $20\ $MHz. A bias magnetic field of $156\ \mu$G is applied
along the light-propagation direction. The data are shown along
with a fit by a dispersive Lorentzian (see Ref. \cite{Mal2004}).
The width of the resonance corresponds to
$\gamma_{exp}^{NMOR}=2\pi\cdot0.7\ $Hz, the narrowest
magneto-optical resonance width observed with alkali atoms to
date.} \label{Fig_FMNMOR_Data}
\end{figure}
An example of a high-resolution recording of the narrow feature is
shown in the lower plot in Fig. \ref{NISTuW2}. The line shape is
well described by a Lorentzian as seen from the fitting curve also
shown in the plot.

We have measured the widths and shifts of the central narrow
resonance for each of the cells. In order to eliminate the effects
of power broadening and shifts, we performed double extrapolation
of the widths and central frequencies of the resonances to zero
optical and microwave power, using their observed linear
dependence on power at low powers. (The effects of light
broadening can also be minimized by judicious choice of light
tuning; for example, the high-frequency slope of the $^{85}$Rb
$F=3\rightarrow F'$ optical transition provides relatively large
signals with greatly reduced light broadening. The signal
deteriorates on the low-frequency slope, presumably because Zeeman
optical pumping dominates over hyperfine pumping. Similar
``tricks" are also used in optical pumping magnetometry
\cite{Ale87} and in NMOR \cite{Bud98,Bud2000Sens}.) The results,
which were found to be independent of laser tuning and
polarization, are summarized in Table I. For $^{85}$Rb,
uncertainties in the shifts include both the errors of the present
measurement and $3\ $Hz uncertainty in the knowledge of the
absolute transition frequency for free atoms \cite{VanierAudoin};
for $^{87}$Rb, the latter uncertainty is negligible \cite{Biz99}.

An example of the data taken with the microwave cavity is shown in
Fig. \ref{NISTuW5}. The TE$_{011}$ cavity has important advantages
over a simple loop, as can be seen from the comparison of Fig.
\ref{NISTuW5} with the middle plot in Fig. \ref{NISTuW2}. In
particular, strong suppression of the broad pedestal and the
$\Delta M_F \ne 0$ microwave transitions is apparent. (In Ref.
\cite{Fru83}, microwave transition lineshapes obtained using
TE$_{011}$ and TE$_{111}$ cavities were compared to each other.
Narrow lines appeared only when the TE$_{011}$ cavity was used.
Ref. \cite{Fru83} also contains references to calculations of the
lineshape in the regime in which the dimensions of the cell are
comparable with the wavelength, and, correspondingly, with the
microwave cavity mode size.)

As an additional cross-check of the results given in Table
\ref{Table1}, a set of room temperature data for the H2 cell was
also taken using the 780$\ $nm D2 resonance (a different diode
laser system was used for this measurement). The results, the
intrinsic width of $20.3(9)\ $Hz and shift of $-89.5(11)\ $Hz, are
consistent with the data obtained with D1 resonance (see Table
\ref{Table1}), as expected.

The data taken at different temperatures in the TT11 cell provide
some insight into the adsorption process of the alkali atoms onto
the walls. The adsorption time of the atom on the wall, under
simplifying assumptions such as a uniform adsorption energy on all
sites on the surface, is usually assumed to be \cite{Gol61}
\begin{equation}
\bar{t}_a=\tau_0 e^{E_{a}/kT},\label{Eq_ads_time}
\end{equation}
where $\tau_0$ is the period of vibration of the adsorbed atom in
the wall potential, $E_{a}$ is the adsorption energy, $k$ is the
Boltzmann constant and $T$ is the absolute temperature. At higher
temperatures, therefore, an atom spends less time on the wall and
should experience a smaller frequency shift and broadening. The
frequency shift (discussed in the following section) in particular
should be reduced by a factor equal to the fractional change in
adsorption time due to the change in cell temperature:
\begin{equation}
\frac{d \Delta \nu}{\Delta \nu} = - \frac{E_{a}}{kT}\cdot
\frac{dT}{T}.
\end{equation}
For the TT11 cell, the frequency shift measured at two
temperatures allows one to calculate [under the assumptions of Eq.
\eqref{Eq_ads_time}] the adsorption energy, giving $E_{a}=0.06\
$eV. This is consistent with previous values found in similar
cells but smaller than the $0.1\ $eV reported by some researchers
(see Ref. \cite{Rah87} and references therein).

In interpreting the observed microwave frequency shifts we have
assumed that these shifts originate in collisions of rubidium
atoms with the cell walls because the cells studied here are
nominally free from buffer gas. However, due to high mobility of
helium atoms in glass, it is possible that atmospheric helium has
diffused into the cells through the glass (particularly in the
case of the oldest Gib cell). Assuming that the helium inside the
cells is in equilibrium with that in the atmosphere (for which the
partial He concentration is $5.2\cdot10^{-6}$) and using the
literature data (e.g., Ref. \cite{Van74}) for the frequency shifts
of the rubidium 0-0 microwave transitions, we find that the
microwave resonance is shifted by $+1.3\ $Hz and $+2.9\ $Hz for
$^{85}$Rb and $^{87}$Rb, respectively. Since these shifts are
relatively small, and since we do not know the actual pressure of
helium in the cells, we ignore these shifts in the rest of the
paper, noting them as a possible source of a small systematic
error. Broadening due to collisions with helium is negligibly
small.

The procedure for obtaining the Zeeman relaxation data is
described in Refs. \cite{Bud2002FM,Mal2004}. Care was exercised to
minimize the possible contributions to resonance linewidths due to
magnetic-field gradients (which were compensated to a level $\sim
1 \mu$G/cm). The contribution of gradient broadening to the FM
NMOR linewidth (upon compensation of the gradients) is found to be
negligible at the level of present uncertainties \cite{Pus2005}.
Similar to the procedure used for the microwave transitions, the
observed resonance widths (see Fig. \ref{Fig_FMNMOR_Data}) were
extrapolated to zero light power.

Atomic number densities listed in Tables \ref{Table1} and
\ref{Table2} were determined by fitting linear absorption spectra
taken at low light power ($\sim\ 1\ \mu$W).

\section{Interpretation}
\label{Sect:Interpretation}

\begin{table*}
\caption{Linewidth budget for the microwave transitions.
$\gamma_{a}^{\mu}/(2\pi)$ and $\gamma_{er}^{\mu}/(2\pi)$ are the
deduced contributions to the microwave linewidth from adiabatic
and electron-randomization collisions (from Tables \ref{Table1}
and \ref{Table2}), respectively. The last two columns list the
total expected width based on the sum of these deduced
contributions and the experimentally observed microwave linewidths
(from Table \ref{Table1}).} \label{Table3}
   \begin{tabular}{c|c|c|c|c|c}
\hline
        Cell & Isotope  & $\gamma_{a}^{\mu}/(2\pi)$ (Hz) & $\gamma_{er}^{\mu}/(2\pi)$ (Hz) & Total (Hz) &  $\gamma_{exp}^{\mu}/(2\pi)$ (Hz) \\
        \hline \hline
         Ale-10 & $^{85}$Rb & 2 & 5 & 7 & 8.7(5) \\
          \hline
         Gib & $^{85}$Rb & 0.6 & 13 & 14 & 11(2) \\
             & $^{87}$Rb & 5 & 9 & 14 & 16(4) \\
          \hline
         H2 & $^{87}$Rb  & 9 & 11 & 20 & 22(3) \\
          \hline
          \hline
    \end{tabular}

\end{table*}
There are several known relaxation mechanisms at work in the vapor
cells studied here, including spin-exchange relaxation, loss of
polarized atoms due to collisions with uncoated surfaces
(primarily in the cell's stem), and relaxation due to collisions
with the wall coating.

The contribution of spin-exchange relaxation to the microwave
width can be estimated using the formulae given in Refs.
\cite{Gro68,Hap72}. When only one isotope with nuclear spin $I$ is
present in the cell, the spin-exchange contribution to the
linewidth of the 0-0 microwave transition is given by
\begin{align}
\frac{\gamma_{se}^{\mu}}{2\pi} =
\frac{\mathcal{R}(I)n\overline{v}_{rel}\sigma_{se}}{\pi},
\label{Eq_SE_Width}
\end{align}
where $\sigma_{se}$ is the spin-exchange-collision cross-section
($\sigma_{se}=2\cdot 10^{-14}\ $cm$^2$ for all cases relevant here
\cite{Hap72}), $n$ is the atomic number density,
\begin{align}
\overline{v}_{rel}=\sqrt{8kT/\pi \mu_{red}}
\end{align}
is the average relative speed, and $\mu_{red}$ is the reduced mass
of the colliding atoms, and
\begin{align}
\mathcal{R}(I)=\frac{6 I +1}{8 I +4}
\end{align}
is the so-called nuclear slow-down factor (see Refs.
\cite{Gro68,Hap72}) for the $0-0$ transition. For the Gib cell,
where two isotopes are present, the considerations are similar.
The values of the contribution of the spin-exchange collisions to
the microwave linewidths deduced from the measured number density
are listed in the next-to-last column of Table \ref{Table1}. In
all cases, this contribution is less than 25\% of the overall
observed linewidth $\gamma^{\mu}_{exp}$.

The relaxation due to the stem can be estimated from the geometry
(except for the Gib cell which does not have a stem, whose
deposits of solid rubidium are apparently covered with paraffin),
and
is found to contribute to the microwave linewidth less than a
fraction of a hertz in the case of the Ale-10 cell and less than
$1\ $Hz in the case of the TT11 and H2 cells. Effects related to
the cell stems therefore contribute to the hyperfine decoherence
at a level somewhat less than the measurement errors given in
Table \ref{Table1}. We ignore this contribution in the following
discussion.

Collisions of alkali atoms with the wall coating can be separated
into three categories. The most ``gentle" or adiabatic collisions,
while causing hyperfine transition frequency shift and decoherence
(as discussed below), generally do not result in population
transfer or Zeeman decoherence. The stronger collisions, for
example, collisions with paramagnetic impurities or ``dangling
bonds," randomize the electron spin. However, they do not affect
the nuclear spin, so a polarized atom retains a certain degree of
polarization after the collision. Finally, an atom can be absorbed
into the coating for a sufficiently long time that all
polarization is destroyed. First, we analyze the contribution of
the adiabatic collisions.

While the adiabatic collisions do not cause atoms to jump between
quantum states, the interaction with the wall during a collision
causes a phase shift $\phi$ between hyperfine states \cite{Gol61}.
The hyperfine transition frequency shift due to this phase shift
can be estimated as follows. Consider an atom in the cell that we
will track for a time $\tau$ much longer than the typical interval
between its wall collisions. The distance travelled by the atom
(along some complicated reticulated trajectory) is
$\overline{v}\tau$, where
\begin{align}
\overline{v}=\sqrt{8kT/\pi M}
\end{align}
is the mean speed
and $M$ is the mass of the atom. The next question
is: how many times did this atom collide with the wall? The answer
for a spherical cell (for which the mean distance between wall
collisions assuming the usual cosine angular distribution of atoms
bouncing off the wall is $4R/3$) is
\begin{align}
\frac{\overline{v}\tau}{4R/3}=\frac{\tau}{t_c}, \label{Eq_for_t_c}
\end{align}
where $t_c$ is the characteristic time between wall collisions. If
the phase shift per collision is $\phi$, the overall phase
increment in time $\tau$ is
\begin{align}
\phi \frac{\overline{v}\tau}{4R/3}. \label{ps_1}
\end{align}
On the other hand, the phase increment is also equal to
\begin{align}
2\pi \Delta \nu  \tau, \label{ps_2}
\end{align}
where $\Delta \nu$ is the frequency shift. Equating \eqref{ps_1}
and \eqref{ps_2}, and cancelling $\tau$, we get
\begin{align}
\Delta \nu = \frac{\phi}{2\pi}\frac{3\overline{v}}{4R}.
\label{EqShift}
\end{align}

The values of the phase shift $\phi$ for various cells,
experimental conditions, and Rb isotopes, deduced from the
experimental values of the microwave transition frequency shifts
using Eq. \eqref{EqShift}, are listed in Table \ref{Table1}. The
results are consistent with the available earlier data for Rb in
alkane-coated cells (Ref. \cite{Rob82} and references therein)
within the spread between different cells under similar conditions
and with the same Rb isotope.

In order to compare the coating properties for cells containing
different isotopes, one can scale the phase shift to one and the
same isotope (e.g., $^{85}$Rb) using the expected proportionality
of the phase shift and the hyperfine transition frequency
\footnote{This neglects possible effects of the different mass of
the two isotopes. The proportionality of the phase shift and the
hyperfine frequency arises from the fact that the mechanism
through which the phase shift occurs is the change of the
valence-electron density near the nucleus during a collision
\cite{Her61,Gol61}. This scaling is confirmed, at least
approximately, by the present work (see the data for the two Rb
isotopes simultaneously present in the Gib cell listed in Table
\ref{Table1}), and the earlier work of Ref. \cite{Van74}.}. The
scaled values ($\phi'$) for the $^{87}$Rb data are listed in Table
\ref{Table1}. The results indicate that all the different coatings
studied in this work produce roughly the same phase shifts in wall
collisions. We also note that the phase shifts per collision
measured here in Rb are roughly consistent with those measured for
Cs on paraffin coatings ($\phi=0.09(1)\ $rad \cite{Gol61}), when
scaled to the corresponding hyperfine frequency.

Because of the statistical character of the collisions, there is a
spread in the amount of phase shift acquired by the atoms, which
contributes to the resonance width \cite{Gol61,Joc81} (a
derivation of the broadening and shift due to this mechanism is
given in the Appendix):
\begin{align}
\frac{\gamma_a^{\mu}}{2\pi} = \frac{\phi^2}{\pi t_c}.
\label{EqWidth}
\end{align}
For the parameters and the frequency shifts measured in the
present experiment, these contributions to the width comprise from
$\gamma_a^{\mu}/(2\pi)\approx 0.6\ $Hz to $\sim 10\ $Hz. As seen
from Table \ref{Table1}, the sums of $\gamma_a^{\mu}$ and
$\gamma_{se}^{\mu}$ are insufficient to explain the observed
overall intrinsic widths $\gamma_{exp}^{\mu}$ in any of the cases.

A possible contribution to the linewidth that may explain this
is from electron-spin randomizing collisions with the wall or,
possibly, gaseous impurities other than helium \cite{Bou64}. The
magnitude of this contribution to the width of the microwave
transitions can be estimated from the assumption that the measured
NMOR linewidths ($\gamma^{NMOR}_{exp}$) are also dominated by
electron-spin randomization collisions \cite{Oku2003}. The
measured NMOR linewidths are broader than what is expected given
the known spin-exchange cross-sections (see Table \ref{Table2}).
The relaxation rate of the 0-0 microwave coherence due to spin
randomization should be 3/4 of the electron randomization rate
\cite{Hap72}, while the intrinsic NMOR linewidth due to
electron-spin randomization is smaller, due to the nuclear
slow-down effect (see, for example, Ref. \cite{Hap72}), and is
calculated to be $\approx 1/3$ of the spin-randomization rate for
$^{85}$Rb and $\approx 1/2$ for $^{87}$Rb \cite{Oku2003}.

Using this information, the contribution of the electron-spin
randomization collisions to the microwave linewidth can be
estimated ($\gamma_{er}^{\mu}$, Table \ref{Table2}). For example,
for the Ale-10 cell at 25$^\circ$C, assuming that the Zeeman
relaxation is dominated by spin-randomization collisions, we have
\begin{align}
\frac{\gamma_{er}^{\mu}}{2\pi} \approx 2\times (1.2\ {\rm
Hz})\times 3 \times \frac{3}{4} \approx 5\ {\rm Hz}.
\label{Eq:ZeeemanRel}
\end{align}
In expression \eqref{Eq:ZeeemanRel}, the factor of two accounts
for the relation between the relaxation rate for the 0-0 coherence
and the Lorentzian width of the microwave transition. (Under the
conditions of our experiments, in which linearly polarized
low-intensity light is used, the effect of spin-exchange
collisions is nearly identical to that of electron-randomization
collisions \cite{Hap72}, and we do not separate spin exchange from
the additional electron-randomization processes in this estimate
and those presented in Tables \ref{Table2} and \ref{Table3}.)
Adding up all the contributions and estimating the associated
uncertainties, we find that for the Ale-10 cell we can account for
about 7(1)$\ $Hz out of the observed 8.7(5)$\ $Hz, which is
satisfactory, particularly, in view of a number of simplifying
assumptions in our model (for example, the assumption that the
dispersion of $\phi$ is equal to $\phi^2$ in the adiabatic
collisions, see Appendix). The microwave linewidth budgets for the
cells where both the microwave and NMOR data are available are
summarized in Table \ref{Table3}.

The fact that both Rb isotopes are simultaneously present in the
Gib cell allows a check of our model for consistency, and provides
further evidence that the dispersion of phase shifts in adiabatic
collisions is not the dominant source of the microwave linewidth.
For the values of $\phi$ (Table \ref{Table1}) for the two isotopes
extracted from the measured frequency shifts using Eq.
\eqref{EqShift}, Eq. \eqref{EqWidth} predicts about an order of
magnitude larger contribution to the width from phase-shift
dispersion in adiabatic collisions for $^{87}$Rb compared to that
for $^{85}$Rb. This is clearly inconsistent with a relatively
small difference in the width observed experimentally, but does
correspond to the prediction of our model.

So far, we have considered adiabatic and electron-spin
randomization collisions. Collisions that completely depolarize
atoms would also contribute to both the hyperfine and Zeeman
relaxation. However, if one assumes that relaxation is dominated
by collisions of this type rather than electron-randomization
collisions, carrying out an analysis similar to the one above, one
does not achieve satisfactory agreement between the microwave and
NMOR linewidths.

The present experimental data for the widths and shifts of the
microwave transitions appears inconsistent with a hypothesis that
the linewidth is dominated by dispersion of the phase shifts in
adiabatic wall collisions. On the other hand, comparing the
microwave data and the Zeeman relaxation data measured with
nonlinear magneto-optical rotation, we have proposed that the
dominant source of the linewidth is electron-spin randomizing
collisions. This hypothesis consistently accounts for the
linewidth-budget deficits for both microwave transitions and NMOR
resonances. The rate of the electron-randomization collisions
necessary to explain the observed microwave and NMOR linewidths is
too large to be accounted for by spin-exchange collisions (Tables
\ref{Table1} and \ref{Table2}). Thus it is necessary to assume
electron randomization processes occurring either in collisions of
the alkali atoms with cell walls or, possibly, in collisions with
gaseous impurities. These two scenarios may be distinguished by
comparing relaxation rates for otherwise similar cells having
vastly different diameters.

It is interesting to compare the present results with those of
Ref. \cite{Van74}. In that work, the rates of the
hyperfine-coherence and population relaxation in wall collisions
in a Paraflint-coated cell were measured, and the latter was found
to be an order of magnitude lower than the former. In the present
experiment, since we have found that electron-randomization
collisions dominate relaxation at room temperature, one would
expect that the two rates should be similar.  This apparent
contradiction may be attributable to an important difference of
the experimental procedure of Ref. \cite{Van74} compared to that
of this work: namely that the data were obtained in the limit of
zero rubidium density in the cell (which was achieved by
maintaining the stem at a temperature lower than that of the cell
walls). This may suggest that the electron randomization in our
case is due to a modification of the coating surface due to the
presence of rubidium atoms. These issues will be addressed in
future work.


\section{Conclusion}

It appears that the vapor-cell coatings studied in this work,
manufactured using three rather different technologies, all show
comparable performance in terms of the parameters (linewidth and
shift) relevant to their use in magnetometers and atomic clocks.
Moreover, the fact that a cell (Gib) manufactured 40 years ago
shows comparable performance to that of more recently manufactured
cells is evidence of the stability of the coating properties
\footnote{For example, assuming that the intrinsic coating
properties at a certain time after manufacturing were identical
for the Ale-10 and the Gib cell, we can roughly estimate the
temporal drift of the microwave frequency shift as being $\le 10\
$Hz/$30\ $years.} and suitability of such cells for extremely long
term measurements, for example, as frequency-reference and
magnetic-sensor elements for deep-space missions.

The durability of the wall coatings and the lack of dependence of
the coating properties on the exact molecular-weight composition
and the method of coating deposition are encouraging with regard
to the application of such coatings to miniature atomic frequency
references. Wall coatings might play an important role in
improving the performance of compact atomic clocks if a way of
integrating the application of the coating with the cell
fabrication process is found. This integration will likely be an
important future step in the development of atomic clocks based on
sub-millimeter alkali-atom vapor cells.

This work was initiated and inspired by H.~G.~Robinson, who has
contributed greatly to its realization. The authors are grateful
to D.~English, S.~M.~Rochester, and J.~E.~Stalnaker for useful
discussions and help with data analysis, to S.~N.~Evans for his
help in understanding the statistics underlying collisional
broadening and shift, and to H.~Shugart, E.~B.~Alexandrov,
M.~V.~Balabas and Symmetricom, TRC for providing the
anti-relaxation-coated cells. This work was supported by the
Office of Naval Research, National Science Foundation (NSF), by a
CalSpace Minigrant, and by the Microsystems Technology Office of
the Defence Advanced Research Projects Agency (DARPA). D.B. also
acknowledges the support of the Miller Institute for Basic
Research in Science.

\section{Appendix}
Here we give a derivation of line broadening and shift arising
from the collisional phase shifts acquired by atoms.

Suppose a collection of atomic oscillators are all in phase
initially, and in the absence of collisions, they all oscillate at
a frequency $\omega_0$. Let $\phi$ be the average phase shift per
collision, and $1/t_c$ the average collision rate. There are two
factors that lead to a dispersion in different atoms' phase shifts
acquired after a time $t\gg t_c$. First, there is a statistical
distribution of the number of collisions $n$ experienced by atoms
over a time $t$ which we will assume to be Poissonian:
\begin{align}
p(n,t)=\frac{e^{-t/t_c}\prn{t/t_c}^n}{n!}, \label{AppEq1}
\end{align}
where $p(n,t)$ is the probability that an atom experiences $n$
collisions in time $t$ [the mean number of collisions
corresponding to the distribution \eqref{AppEq1} is $\langle n
\rangle = t/t_c$]. Second, the phase shifts per collision are not
the same. We will assume a normal distribution with a mean value
$\phi$ ($|\phi|\ll 1$) and a dispersion $\phi^2$. (The latter
property follows from a distribution of wall-sticking times with a
universal binding energy exceeding $kT$; see, for example, Ref.
\cite{Gol61}.)

Let us consider atoms that have experienced some fixed number of
collisions $n\gg 1$. Let $\phi_n$ be the overall phase accumulated
by an atom over $n$ collisions. Because of the normal distribution
of phase shifts in individual collisions (resulting in a random
walk in phase), we have a Gaussian distribution of accumulated
phases:
\begin{align}
p(\phi_n,n)=\frac{1}{\sqrt{2\pi
n\phi^2}}e^{-\frac{\prn{\phi_n-n\phi}^2}{2n\phi^2}},
\label{AppEq2}
\end{align}
where $n\phi$ is the average phase accumulated in $n$ collisions,
and $n\phi^2$ is the dispersion.

Taking into account the distributions \eqref{AppEq1} and
\eqref{AppEq2}, the oscillation amplitude averaged over the atomic
ensemble is found as a weighted sum of the contributions from
individual atoms ($\propto e^{i (\omega_0 t + \phi_{j})}$, where
$\phi_{j}$ is the phase accumulated by this individual atom):
\begin{align}
&A(t)\propto \nonumber
\\&\sum_{n=0}^{\infty}\frac{e^{-t/t_c}\prn{t/t_c}^n}{n!}
\int_{-\infty}^{\infty}\frac{e^{i (\omega_0 t +
\phi_n)}}{\sqrt{2\pi
n\phi^2}}e^{-\frac{\prn{\phi_n-n\phi}^2}{2n\phi^2}}d\phi_n\\
&=e^{i\omega_0 t-t/t_c\prn{1-e^{-i \phi-\phi^2/2}}},
\label{AppEq3}
\end{align}
where in the last step we have explicitly evaluated the integral
and the sum. Next, we use the fact that $|\phi|\ll 1$, and,
expanding the exponential factor to second order in $\phi$, we
obtain
\begin{align}
A(t)\propto e^{i(\omega_0-\phi/t_c)t-\phi^2 t/t_c}, \label{AppEq4}
\end{align}
which says that the frequency of the oscillation is shifted by
$\phi/t_c$, and the amplitude decays at a rate $\phi^2/t_c$,
leading to line broadening. (It is interesting to note that
neglecting either one of the random factors -- the number of
collisions experienced by an atom or the dispersion in phase shift
per collision -- leads to a two times slower decay rate in either
case.) In order to obtain the line width of the absorption
resonance, the decay rate of the amplitude has to be multiplied by
a factor of two. Therefore, Eq. \eqref{AppEq4} gives the formulae
\eqref{EqShift} and \eqref{EqWidth}.

\bibliography{NMObibl}

\end{document}